\documentclass[fleqn,usenatbib]{mnras}

\usepackage[T1]{fontenc}
\usepackage{ae,aecompl}

\usepackage{graphicx}   
\usepackage{amsmath}    
\usepackage{amssymb}    

\usepackage{txfonts}
\usepackage{hyperref}
\usepackage{breakurl}

\title[The double degenerate system NLTT 12758]{A fast spinning
magnetic white dwarf in the double-degenerate, super-Chandrasekhar
system NLTT 12758 \thanks{Based on observations made with ESO
telescopes at the La Silla Paranal Observatory under programmes
083.D-0540, 084.D-0862, 089.D-0864 and 090.D-0473. Based in part
on data collected with the Danish 1.54-m telescope at the ESO La
Silla Observatory.}}

\author[A. Kawka et al.]{Adela Kawka$^{1}$\thanks{E-mail:
kawka@asu.cas.cz}, Gordon P. Briggs$^{2}$, St\'ephane Vennes$^{1,3}$, 
Lilia Ferrario$^2$
\newauthor
Ernst Paunzen$^{4}$ and Dayal T. Wickramasinghe$^2$\\ \\
$^{1}$Astronomick\'y \'ustav AV \v{C}R, Fri\v{c}ova 298, 251 65 Ond\v{r}ejov,
Czech Republic\\
$^{2}$Mathematical Sciences Institute, The Australian National University, 
Canberra, ACT 0200, Australia\\
$^{3}$Visitor at Mathematical Sciences Institute, The Australian National University, 
Canberra, ACT 0200, Australia\\
$^{4}$Department of Theoretical Physics and Astrophysics, Masaryk University,
Kotl\'a\v{r}sk\'a\ 2, 611 37 Brno, Czech Republic}

\date{Accepted XXX. Received YYY; in original form ZZZ}

\pagerange{\pageref{firstpage}--\pageref{lastpage}} 

\pubyear{2015}

\begin{document}
\label{firstpage}
\pagerange{\pageref{firstpage}--\pageref{lastpage}}
\maketitle

\begin{abstract}
We present an analysis of the close double degenerate NLTT\,12758,
which is comprised of a magnetic white dwarf with a field of about
3.1 MG and an apparently non-magnetic white dwarf. We measured an orbital period
of 1.154 days and found that the magnetic white dwarf is spinning
around its axis with a period of 23 minutes. An analysis of the
atmospheric parameters has revealed that the cooling ages of the two
white dwarfs are comparable, suggesting that they formed within a
short period of time from each other. Our modelling indicates that
the non-magnetic white dwarf is more massive ($M=0.83\, M_\odot$) than
its magnetic companion ($M=0.69\,M_\odot$) and that the total
mass of the system is higher than the Chandrasekhar mass. Although the 
stars will not come into contact over a Hubble time, when they
do come into contact, dynamically unstable mass
transfer will take place leading to either an accretion induced
collapse into a rapidly spinning neutron star or a Type Ia
supernova.
\end{abstract}


\section{Introduction}

The majority of stars will evolve into a white dwarf and a significant
fraction of white dwarfs harbours a magnetic field that ranges from a
few kG to about 1000 MG \citep{lie2003, kaw2007}. Spectroscopic and
spectropolarimetric surveys
\citep[e.g.,][]{sch1995,sch2001a,azn2004,kaw2007,kaw2012,lan2012,kep2013}
of white dwarfs have been able to place constraints on the incidence
of magnetism among white dwarfs. The incidence of magnetic white
dwarfs in the local neighbourhood has been estimated by
\citet{kaw2007} to be around 20\,\%. The local sample, as well as
various surveys, have shown that the incidence of magnetism as a
function of field strength is constant, although \citet{lan2012}
suggested a possible field resurgence at the extremely low-field ($<1$
kG) end of the distribution. A higher incidence of magnetism is also
observed in cool polluted white dwarfs. \citet{kaw2014} found an
incidence of $\approx 40$\,\% in cool ($T_{\rm eff} < 6000$ K)
DAZ\footnote{DAZ type white dwarfs show photospheric hydrogen (DA) and metal lines.}
white dwarfs. A higher incidence of magnetism was also observed among
cool DZ\footnote{DZ type white dwarfs show metal lines only.} white dwarfs
\citep{hol2015}. A recent review on the properties of magnetic white
dwarfs can be found in \citet{fer2015a}.

The origin of large scale magnetic fields in stars is still one of the
main unanswered questions in astrophysics, although recent data,
particularly from surveys such as the Sloan Digital Sky Survey
\citep[SDSS, ][]{York2000}, the Magnetism in Massive Stars
\citep[MiMes, ][]{wad2016} and the Binarity and Magnetic Interactions
in various classes of stars \citep[BinaMIcS, ][]{ale2015} may have
finally thrown some light into this matter \citep{fer2015b}. Magnetism
in white dwarfs has been explained with two main evolutionary
scenarios. For a long time the leading theory was that the progenitors
of magnetic white dwarfs are magnetic Ap and Bp stars
\citep{ang1981}. Under the assumption of magnetic flux conservation,
the magnetic field strengths observed in Ap stars would correspond to
magnetic fields in white dwarfs in excess of 10 MG
\citep{kaw2004,tou2004,wic2005}. The progenitors of white dwarfs
with weaker fields may be other main-sequence stars whose magnetic
fields are below our current detection limits or could be
dynamo-generated in later stages of stellar evolution.

More recently, proposals that strong magnetic fields are created in
evolving interacting binaries via a dynamo mechanism during a 
common envelope (CE) phase \citep{tou2008,pot2010,nor2011,gar2012,wic2014}
have gained momentum as a possible origin for strong magnetic fields
in white dwarfs. The main reason for this proposal is that all
magnetic white dwarfs appear to be either single or in interacting
binaries (the magnetic cataclysmic variables). That is,
magnetic white dwarfs are never found paired with a non-interacting,
non-degenerate star, which is at odds with the fact that approximately
30\% of all non-magnetic white dwarfs are found in non-interacting
binaries with a non-degenerate companion (usually an M-dwarf)
\citep{lie2005,fer2012,lie2015}. This result is hard to explain and
leaves the magnetic cataclysmic variables without obvious
progenitors. Because of this observational peculiarity, the existence
of magnetic fields in white dwarfs has been linked to fields generated
during CE binary interactions or mergers. The merger scenario during
the CE also successfully explains the higher than average mass of
isolated magnetic white dwarfs \citep{bri2015}. The complex magnetic
field structure usually observed in rotating high field magnetic white dwarfs
would also be in support of a merging hypothesis.

However, a few common-proper motion (CPM) magnetic plus non-magnetic
double degenerate systems are now known
\citep{fer1997,gir2010,dob2012,dob2013}. In some of these cases, the
more massive magnetic white dwarf is hotter and hence younger than its
non-magnetic companion, which seems to imply that the more massive
star evolved later.  This apparent paradox can be resolved by
postulating that systems of this kind were initially triple
systems and that the magnetic white dwarf resulted from the merger of
two of the three stars \citep[e.g., EUVE\,J0317$-$855, ][]{fer1997}.

The study of the magnetic field structure in white dwarfs may also
give us important clues on how they formed. Normally, a simple dipole
is assumed for the field structure, but the study of rotating magnetic
white dwarfs have all shown variability, hence revealing much more
complex structures. One of the most extreme examples of a rotating
magnetic white dwarf is the hot ($T \approx 34\,000$ K) and massive
($M\approx 1.35\ M_\odot$) EUVE\,J0317-855, which has a rotation
period of 12 minutes \citep{bar1995,fer1997}. The rotation of the
white dwarf reveals a two component magnetic field structure: A high
field magnetic spot ($B \ge 425$ MG) with an underlying lower field
\citep{ven2003}. Another example of a rotating white dwarf with a
complex magnetic field structure is WD\,1953-011
\citep{max2000,val2008}.  In this case, the rotation is slower
\citep[$P_{\rm rot} = 1.448$ days][]{bri2005} and the magnetic field
strength is much weaker (180 kG - 520 kG) than that of
EUVE\,J0317-855.

NLTT\,12758 was discovered to be a magnetic white dwarf by
\citet{kaw2012}.  They showed that the circular polarization spectra
are variable and that there is also variability in the H$\alpha$ core
suggesting that NLTT\,12758 is a close double degenerate system. Here,
we present our analysis of spectroscopic, spectropolarimetric and
photometric data of NLTT\,12758. The observations are presented in
Section 2.  The orbital and rotation period analyses are described in
Sections 3.1 and 3.2, respectively. The stellar and atmospheric
parameters are presented in Section 3.3, and we discuss the
evolutionary scenarios in Section 3.4. We discuss the case of
NLTT\,12758 in comparison to other known double degenerate systems
containing a magnetic white dwarf in Section 4 and we conclude in
Section 5.

\section{Observations}

\subsection{Spectroscopy and Spectropolarimetry}

\begin{figure}
\includegraphics[width=1.0\columnwidth]{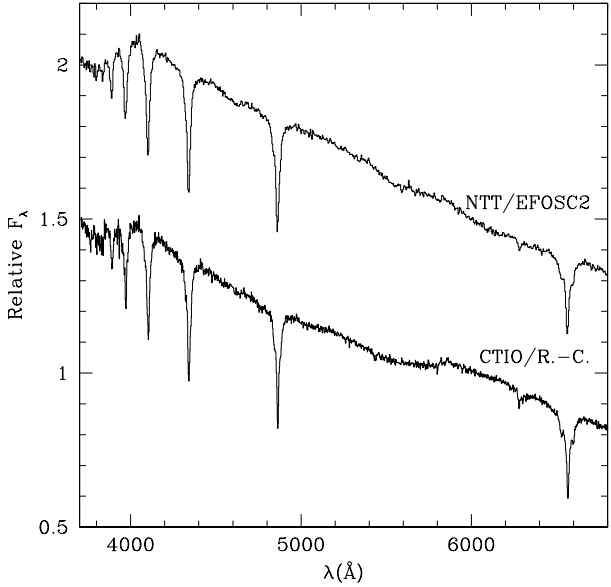}
\caption{Low dispersion CTIO/R.-C. and NTT/EFOSC2 spectra of NLTT\,12758 revealing
Zeeman splitted Balmer lines.
\label{fig_efosc2}}
\end{figure}

NLTT\,12758 was first observed with the R.-C. spectrograph attached to
the 4m telescope at Cerro Tololo Inter-American Observatory (CTIO) on
UT 2008 February 24. We used the KPGL2 (316 lines per mm) grating with
the slit-width set to 1.5 arcsec providing a resolution of about 8
\AA. We obtained a second set of low-dispersion spectra with the
EFOSC2 spectrograph attached to the New Technology Telescope (NTT) at
La Silla. Two consecutive spectra were obtained on UT 2009 08 27. We
used grism number 11 and set the slit-width to 1.0 arcsec providing a
resolution of about 14 \AA. Both sets of spectra revealed Zeeman
splitting in the Balmer lines. Figure~\ref{fig_efosc2} shows the low
dispersion spectra.

We obtained a first set of spectropolarimetric observations using the
FOcal Reducer and low dispersion Spectrograph (FORS2) attached to the 8m
telescope (UT1) of the European Southern Observatory (ESO) in 2009. We
obtained another set of observations using the same set-up in 2013. We used
the 1200 lines mm$^{-1}$ grism (1200R+93) centred on H$\alpha$ providing a
spectral dispersion of 0.73 \AA\ pixel$^{-1}$. We set the slit-width to
1 arcsec providing a spectral resolution of 3.0 \AA. Each spectropolarimetric
observation consisted of two individual exposures, with the first having the
Wollaston prism rotated to $-45^\circ$ immediately followed by the second
exposure with the Wollaston prism rotated to $+45^\circ$.

We also obtained five spectra of NLTT\,12758 with the EFOSC2 spectrograph
in September 2012. These spectra were obtained with grism number 20 which
provides a spectral dispersion of 1.09 \AA\ per binned pixel. The slit-width
was set to 0.7 arcsec providing a resolution of 3.0 \AA.

Finally we obtained a set of five consecutive spectra of NLTT\,12758 with the
X-shooter spectrograph \citep{ver2011} attached to the VLT at Paranal 
Observatory on UT 2014 August 26. The spectra were obtained with the slit width
set to 0.5, 0.9 and 0.6 arcsec for the UVB, VIS and NIR arms, respectively. 
This setup provided a resolution of $R=9000$, 7450 and 7780 for the UVB,
VIS and NIR arms, respectively.

The log of the spectroscopic observations is presented in Table~\ref{tbl_log}.

\begin{table}
\centering
\caption{Spectroscopic observation log \label{tbl_log}}
\begin{tabular}{llcc}
\hline
UT date & UT start & Exposure time (s) & Instrument/Telescope \\
\hline
24 Feb 2008 & 02:04:42 & 1200 & RC/CTI04m \\
24 Feb 2008 & 02:26:56 & 1200 & RC/CTI04m \\
27 Aug 2009 & 09:39:21 & 600 & EFOSC2/NTT \\
27 Aug 2009 & 09:49:57 & 600 & EFOSC2/NTT \\
23 Oct 2009 & 06:35:09 & 900 & FORS2/UT1 \\
23 Oct 2009 & 06:51:18 & 900 & FORS2/UT1 \\
23 Oct 2009 & 07:16:20 & 900 & FORS2/UT1 \\
23 Oct 2009 & 07:32:28 & 900 & FORS2/UT1 \\
23 Oct 2009 & 07:55:18 & 900 & FORS2/UT1 \\
23 Oct 2009 & 08:11:35 & 900 & FORS2/UT1 \\
24 Nov 2009 & 02:40:00 & 900 & FORS2/UT1 \\
24 Nov 2009 & 02:56:08 & 900 & FORS2/UT1 \\
24 Nov 2009 & 03:24:14 & 900 & FORS2/UT1 \\
24 Nov 2009 & 03:40:22 & 900 & FORS2/UT1 \\
02 Sep 2012 & 08:41:00 & 900 & EFOSC2/NTT \\
02 Sep 2012 & 09:04:14 & 900 & EFOSC2/NTT \\
03 Sep 2012 & 08:09:10 & 900 & EFOSC2/NTT \\
03 Sep 2012 & 08:33:13 & 900 & EFOSC2/NTT \\
03 Sep 2012 & 09:18:43 & 900 & EFOSC2/NTT \\
04 Jan 2013 & 03:49:14 & 700 & FORS2/UT1 \\
04 Jan 2013 & 04:02:02 & 700 & FORS2/UT1 \\
04 Jan 2013 & 04:15:06 & 700 & FORS2/UT1 \\
04 Jan 2013 & 04:27:55 & 700 & FORS2/UT1 \\
04 Jan 2013 & 04:41:01 & 700 & FORS2/UT1 \\
04 Jan 2013 & 04:53:49 & 700 & FORS2/UT1 \\
07 Jan 2013 & 02:39:34 & 700 & FORS2/UT1 \\
07 Jan 2013 & 02:52:22 & 700 & FORS2/UT1 \\
07 Jan 2013 & 03:05:26 & 700 & FORS2/UT1 \\
07 Jan 2013 & 03:18:14 & 700 & FORS2/UT1 \\
07 Jan 2013 & 03:31:18 & 700 & FORS2/UT1 \\
07 Jan 2013 & 03:44:06 & 700 & FORS2/UT1 \\
07 Jan 2013 & 03:57:28 & 700 & FORS2/UT1 \\
07 Jan 2013 & 04:10:17 & 700 & FORS2/UT1 \\
07 Jan 2013 & 04:33:02 & 700 & FORS2/UT1 \\
07 Jan 2013 & 04:46:02 & 700 & FORS2/UT1 \\
03 Feb 2013 & 03:01:49 & 700 & FORS2/UT1 \\
03 Feb 2013 & 03:14:38 & 700 & FORS2/UT1 \\
03 Feb 2013 & 03:27:35 & 700 & FORS2/UT1 \\
03 Feb 2013 & 03:40:24 & 700 & FORS2/UT1 \\
03 Feb 2013 & 03:53:22 & 700 & FORS2/UT1 \\
03 Feb 2013 & 04:06:10 & 700 & FORS2/UT1 \\
26 Aug 2014 & 08:18:30 & 450/540$^1$ & Xshooter/UT3 \\
26 Aug 2014 & 08:28:43 & 450/540$^1$ & Xshooter/UT3 \\
26 Aug 2014 & 08:37:51 & 450/540$^1$ & Xshooter/UT3 \\
26 Aug 2014 & 08:47:00 & 450/540$^1$ & Xshooter/UT3 \\
26 Aug 2014 & 08:56:08 & 450/540$^1$ & Xshooter/UT3 \\
\hline
\end{tabular}\\
$^1$ Exposure times for the VIS/UVB arms, respectively.
\end{table}

\subsection{Photometry}

We collected available photometric measurements from the Galaxy 
Evolutionary Explorer ($GALEX$) sky survey, optical photometry from 
\citet{egg1968} and the AAVSO Photometric All-Sky Survey, Deep Near Infrared 
Survey (DENIS) of the southern sky, 
the Two Micron All Sky Survey (2MASS) and the Wide-field Infrared Survey
Explorer ($WISE$). These measurements are listed in Table~\ref{tbl_phot}.

\begin{table}
\centering
\caption{Photometric measurements of NLTT\,12758 \label{tbl_phot}}
\begin{tabular}{lcc}
\hline
Band & Magnitude & Reference \\
\hline
$GALEX$ $FUV$ & not detected     & 1 \\
$GALEX$ $NUV$ & $17.401\pm0.016$ & 1 \\
$V$       & 15.46, $15.483\pm0.071$ & 2,3 \\
$B-V$     & $+0.31$          & 2 \\
$U-B$     & $-0.71$          & 2 \\
$B$       & $15.855\pm0.094$ & 3 \\
$g$       & $15.607\pm0.037$ & 3 \\
$r$       & $15.417\pm0.074$ & 3 \\
$i$       & $15.443\pm0.132$ & 3 \\
DENIS $I$   & $14.976\pm0.07$  & 4 \\
DENIS $J$   & $14.713\pm0.15$  & 4 \\
2MASS $J$   & $14.809\pm0.032$ & 5 \\
2MASS $H$   & $14.723\pm0.071$ & 5 \\
2MASS $K$   & $14.683\pm0.096$ & 5 \\
$WISE$ $W1$ & $14.703\pm0.034$ & 6 \\
$WISE$ $W2$ & $14.781\pm0.069$ & 6 \\
\hline
\end{tabular}\\
References: (1) \citet{mor2007}; (2) \citet{egg1968}; (3) \citet{hen2016}; 
(4) \citet{fou2000}; (5) \citet{skr2006}; (6) \citet{cut2012}
\end{table}

We obtained new CCD photometric measurements with the 1.54-m Danish telescope
at the La Silla Observatory in Chile on UT 26th December 2014, 30th January
2015 and 11th March 2016. On 26th December 2014, we alternated between the 
$V$ and $R$ filter and on 30th January 2015 11th March 2016 we observed 
NLTT\,12758 with the $R$
filter only.  The integration time was set to 40 seconds for all
observations. The data reduction and differential photometry were
performed using the C-Munipack
package\footnote{http://c-munipack.sourceforge.net/}. Since several
comparison stars were available, and these were checked individually to
exclude variable objects.  We compared the results of the final
differential light curves using the aperture photometry routine from
IRAF \citep{Stet87}. We found no differences above the photon noise.

\begin{figure*}
\includegraphics[width=0.8\textwidth]{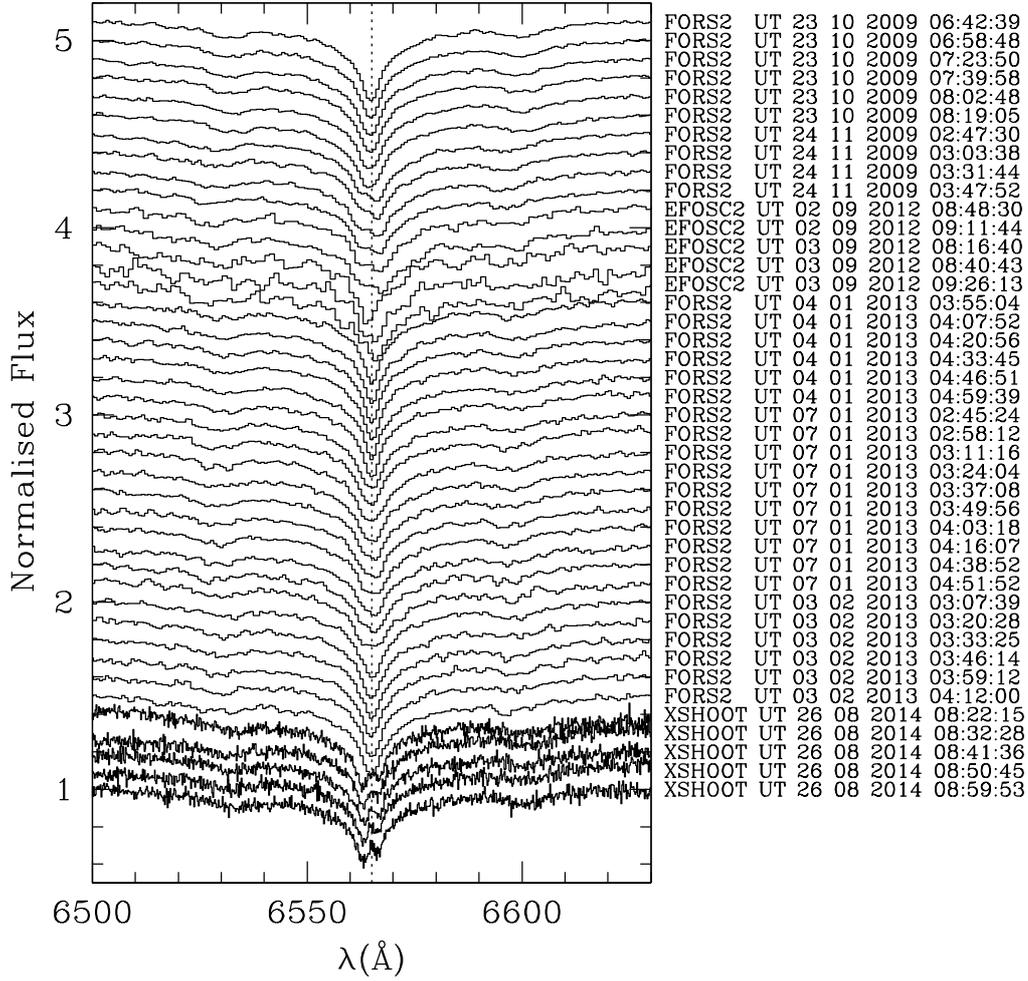}
\caption{EFOSC2, FORS2 and X-shooter spectra of NLTT\,12758 showing variations 
in the H$\alpha$ core. The mid-exposure UT time is listed for each spectrum.}
\label{fig_rad}
\end{figure*}

\section{Analysis}

During the first spectropolarimetric observations of NLTT\,12758, we
found that the $\sigma$ components of H$\alpha$ varied with a reversal
in the polarisation spectra, thus revealing itself as a new member of
the DAP white dwarf class\footnote{DAP white dwarfs show hydrogen
  lines with detectable polarisation. The DAH classification is
  reserved for Zeeman splitted line spectra, but without confirmed
  polarization.}. We also found that the width of the core of the
$\pi$ component is structured and variable, thus suggesting the
presence of a close companion. The FORS2, EFOSC2 and X-shooter spectra
displayed in Figure~\ref{fig_rad} clearly show the variations in the
central H$\alpha$ core. The resolution of the X-shooter spectra and
timing of the observations allowed us to discern the individual cores
of the two components.

\subsection{Binary parameters}

We measured the radial velocity of the magnetic white dwarf by first
subtracting a template representing the DA white dwarf and then
cross-correlating the DAP white dwarf FORS/EFOSC2 spectra ($\sigma$
components only) with the X-shooter spectrum.  The DA radial velocity
could only be measured at quadrature, i.e, at maximum line core
separation, and with a sufficient signal-to-noise ratio. Only three
sets of spectra met these criteria.  Consecutive exposures (2 to 4)
were co-added to increase the signal-to-noise and improve the
reliability of the velocity measurements while minimizing orbital
smearing.  Table~\ref{tbl_rad_vel} lists the barycentric julian date
(BJD) with the measured radial velocities of the magnetic and
non-magnetic white dwarfs in NLTT\,12758. All velocities are
barycentric corrected.

\begin{table}
\centering
\caption{Radial velocity measurements\label{tbl_rad_vel}}
\begin{tabular}{lcc}
\hline
BJD ($2450000+$) & $\varv_{DAP}$ (km~s$^{-1}$) & $\varv_{DA}$ (km~s$^{-1}$) \\
\hline
5127.78952 &  $157\pm6$ & ... \\
5127.81811 &  $165\pm5$ & ... \\
5127.84523 &  $196\pm5$ & ... \\
5159.62676 &    $2\pm8$ & ... \\
5159.64212 &   ...      & $156\pm20$ \\
5159.65748 &   $18\pm8$ & ... \\
6172.87606 &   $18\pm7$ & $196\pm20$ \\
6173.86764 &   $35\pm8$ & ... \\
6296.67107 &  $112\pm6$ & ... \\
6296.68904 &  $120\pm7$ & ... \\
6296.70703 &  $124\pm6$ & ... \\
6299.62701 &   $47\pm7$ & ... \\
6299.65848 &   $38\pm8$ & ... \\
6299.69457 &   $29\pm9$ & ... \\
6326.64042 &  $100\pm6$ & ... \\
6326.66725 &   $79\pm6$ & ... \\
6895.86223 &  $193\pm5$ & $12\pm5$ \\
\hline
\end{tabular}\\
\end{table}

We searched for a period in the measurements using $\chi^2$
minimization techniques by fitting the sinusoidal function
$v=\gamma + K\times \sin{(2\pi(t-T_0)/P)}$ to the measured radial
velocities where $t$ is time (BJD).  The initial epoch ($T_0$), period
($P$), mean velocity ($\gamma$) and velocity semi-amplitude ($K$) were
determined simultaneously and we normalized the $\chi^2$ function by
setting the minimum reduced $\chi^2$ to 1.

\begin{figure}
\includegraphics[width=1.0\columnwidth]{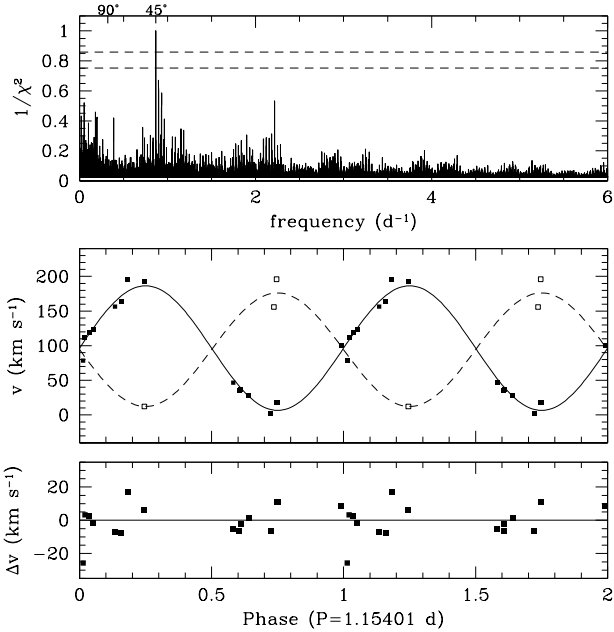}
\caption{(Top panel) period analysis of the FORS2, EFOSC2 and
X-shooter data with 66 and 90\% confidence level (dashed
lines). (Middle panel) radial velocity measurements
(Table~\ref{tbl_rad_vel}) of the DA (open squares) and DAP stars
(full squares) phased on the orbital period and the best-fitting
sine curves (Table~\ref{tbl_param}) and (bottom panel) velocity
residuals for the DAP star. The longest period is marked at $90^\circ$ on
the top horizontal axis along with the actual period at $45^\circ$.}
\label{fig_period}
\end{figure}

Figure~\ref{fig_period} shows the period analysis of the FORS2, EFOSC2
and X-shooter data sets and Table~\ref{tbl_param} lists the new binary 
parameters. Using the FORS2 and EFOSC2 data combined
with the X-shooter data we determined a period of $1.15401\pm0.00005$
days and a velocity semi-amplitude for the DAP star of
$89.7\pm3.8$\,km\,s$^{-1}$ with an average residual of only
7.7\,km\,s$^{-1}$ and commensurate with measurement errors
(Table~\ref{tbl_rad_vel}). The corresponding mass function is
$f(M_{\rm DA}) = 0.0863\pm0.0110\ M_\odot$. Since the X-shooter
spectra were taken near quadrature and clearly show the cores of both
components, we were able to estimate a semi-amplitude of
$81.9\pm17.3$\,km\,s$^{-1}$ for the non-magnetic white dwarf. The orbital mass
ratio $M_{\rm DA}/M_{\rm DAP}=0.85-1.35$ is not sufficiently accurate
to constrain the evolution of the system, and additional constraints will
be provided by the spectroscopic analysis (Section~\ref{AtmParam}).

\subsection{Rotation}

The spectropolarimetric data have revealed a modulation that we
attribute to the rotation of the magnetic white dwarf.

\begin{figure}
\includegraphics[width=1.0\columnwidth]{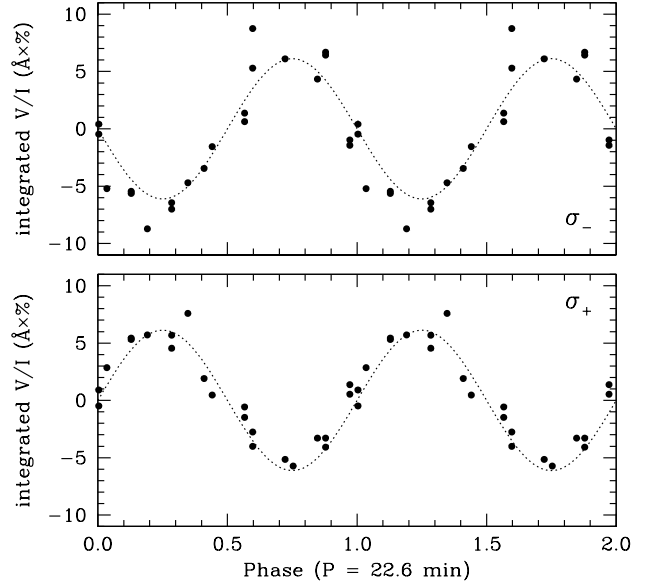}
\caption{Integrated polarization measurements of the two individual
$\sigma$
  components phased on the rotation period of 22.6 minutes revealing a
  complete reversal of the field vector. The top panel shows the
  measurements for the blue-shifted $\sigma_{-}$ component and the bottom
  panel shows the measurements of the red-shifted $\sigma_{+}$
  component.}
\label{fig_rot_per}
\end{figure}

We measured the integrated polarization for both $\sigma$ components
and conducted a period search. Two significant periods at 22.6 minutes
and 9 minutes stand out. Since some of the exposure times were longer
than 9 minutes, it is unlikely that the 9 minutes period is
real. Figure~\ref{fig_rot_per} shows line polarization
measurements obtained by integrating $V/I$ over the wavelength range
($\approx \pm20$\AA) covered
by the individual
$\sigma$ components phased on the 22.6 minute period. Both $\sigma$
components show sinusoidal behaviour and a symmetry about the null
polarization axis which imply that the magnetic poles spend nearly
equal time in the field-of-view.

\begin{figure}
\includegraphics[width=1.0\columnwidth]{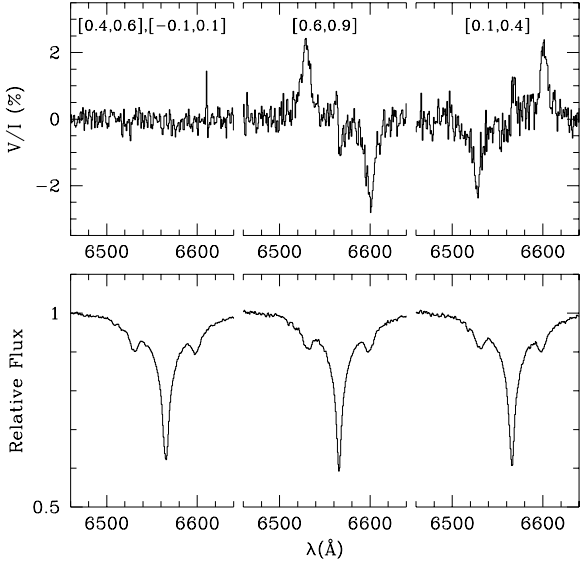}
\caption{Co-added FORS2 circular polarization spectra (top panel) and
  flux spectra (bottom panel) at three phase ranges showing the flip
  in the sign of the $\sigma$ components of H$\alpha$. The spectrum
  with zero polarization corresponds to a nearly orthogonal viewing
  angle to the magnetic axis.}
\label{fig_forsx}
\end{figure}

Figure~\ref{fig_forsx} shows the co-added FORS2 circular polarization
spectra over three separate ranges of a rotation cycle ($P=22.6$\,min)
highlighting the flipping of the sigma components.  The flip in the
sign of the H$\alpha$ $\sigma$ components at phases 0.1-0.4 and
0.6-0.9 and their anti-symmetric behaviour around the zero
polarization spectrum of phases 0.4-0.6, indicate that the magnetic
axis must be nearly perpendicular to the rotation axis of the white
dwarf.

\begin{figure}
\includegraphics[width=1.0\columnwidth]{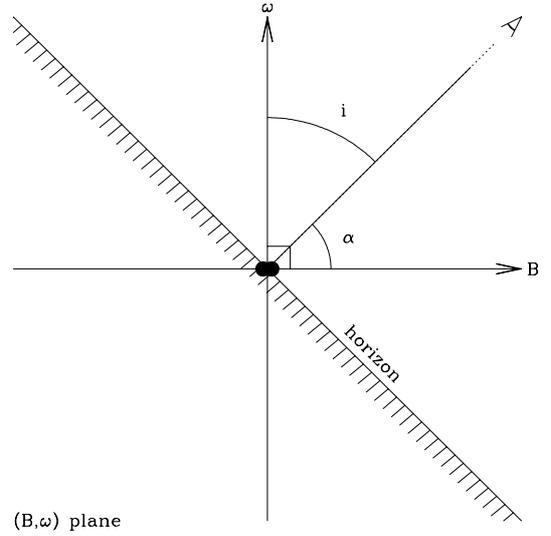}
\caption{Schematic view of the geometry of the double degenerate
  system NLTT\,12758. The rotation plane of the magnetic white dwarf
  is assumed to coincide with the orbital plane, and the spin axis is
  marked $\omega$. The spin axis is at an angle $i$ with respect to
  the observer and the magnetic field axis $B$ is at an angle $\alpha$
  with respect to the observer.}
\label{fig_abs}
\end{figure}

Figure~\ref{fig_abs} illustrates the geometry of the system with
$\alpha$ set at its minimum value ($90-i$).  Assuming $i=45^\circ$
(see Section \ref{AtmParam}), the angle $\alpha$ will vary between
$90^\circ-45^\circ$ and $90^\circ+45^\circ$. When
$\alpha \approx 90^\circ$, the positive and negative polarization
contributions cancel each other and give rise to the unpolarized,
featureless spectrum observed in the phase range 0.4-0.6.  This can be
explained by the change, due to stellar rotation, between the magnetic
field direction and the line of sight to the observer averaged over
the visible hemisphere of the star \citep{wic1979}. The $\pi$
component in the circular polarisation spectra shows the presence of
narrow antisymmetric circular polarisation features. These are caused
by Faraday mixing due to magneto-optical effects which converts linear
polarisation into circular polarisation \citep{mar1981,mar1982} during
the radiation transfer.

\subsubsection{Photometric variations}

The photometric observations were analysed using three different
methods described in detail by \citet{pau2016}.  First, we employed
periodic orthogonal polynomials which are particularly useful for the
detection of non-sinusoidal oscillations. We fitted the observations
to identify the period and employed the analysis of the variance
(ANOVA) statistic to evaluate the quality of the fit \citep{sch1996}.

Next, we employed the string-length methods which simply minimize the
separation between consecutive phased data points at trial periods.
The best-fitting period corresponds to a minimum in the
``string-length'' which consists of the sum of data separations.  The
methods are useful for sparse data sets.

Finally, The Phase Dispersion Minimization (PDM) method is similar to
the string-length method \citep{ste1978}.  In this method, the data
are sorted into phase bins at trial periods and the variance within
each bin is calculated. The sum of the variances is minimized at the
best-fitting period.

\begin{figure}
\includegraphics[width=1.0\columnwidth]{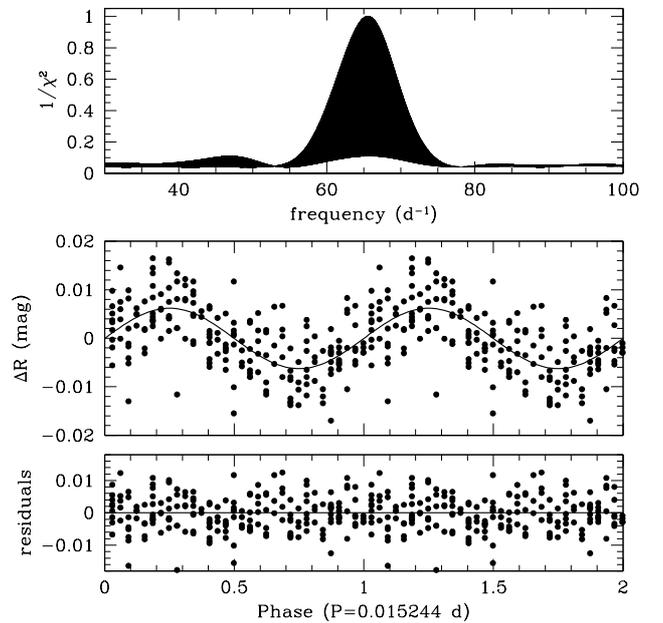}
\caption{(Top panel) period analysis of the measured $R$ photometric
  measurements.  (Middle panel) photometric $R$ magnitudes phased on
  the best rotation period and (bottom panel) residuals.}
\label{fig_phot}
\end{figure}

We found that the photometric observations in the $R$ band show
variations. The calculated frequencies and their errors for the three different 
nights are 65.4\,$\pm$\,1.3, 65.3\,$\pm$\,0.6, and 65.6\,$\pm$\,1.2\, cycles
per day, respectively. The errors depend on the individual data set
lengths and the overall quality of the nights. Within the errors,
these values transform to a period of 22\,$\pm$\,0.5\,min. The semi-amplitude
of the variations is 6.2 mmag.  Figure~\ref{fig_phot} shows the photometric 
magnitudes phased on the best rotation period of 22.0 minutes with the 
periodogram.

We conclude that the variations in spectropolarimetry and photometry
coincide and are phased on the rotation period of the magnetic white
dwarf. The photometric variation may be explained in terms of
magnetic dichroism which is caused by the different absorption
coefficients of left and right handed circularly polarised
radiation. A formulation for magnetic dichroism of hydrogen in
magnetic white dwarfs was first obtained by \citet{lam1974} and used
to explain the photometric variations of the high field magnetic white dwarf
EUVE\,J0317-855 \citep{fer1997}. However the magnetic field of the DAP
component of NLTT\,12758 is relatively low ($B < 20$ MG) for this
effect to be important. An alternate explanation for the photometric 
variations could be stellar spots \citep{bri2005}. Such a spot could be formed
by the inhibition of convection in the atmosphere by the magnetic field.
\citet{tre2015} show that convection is inhibited at the surface of objects
such as the magnetic component of NLTT~12758, however
their models are not able to explain flux variations like those observed 
in NLTT~12758 and other cool white dwarfs with low magnetic fields observed by
\citet{bri2013}.

\subsection{Stellar and atmospheric parameters}

\subsubsection{Modelling the field structure}

The appearance of the spectra of magnetic white dwarfs changes
dramatically as the field increases in strength.  If we indicate with
$(n,l,m_l)$ the zero field quantum numbers, the linear Zeeman regime
arises through the removal of the $m_l$ degeneracy, which for the
Balmer series occurs at a field strength of $\sim 1-4$\,MG.  As the
field increases, or the principal quantum number $n$ increases, the
quadratic effect becomes more important until the $l$ degeneracy is
also removed. This is called the quadratic Zeeman regime. In this
regime, the wavelength shift depends on the electron excitation level
and the $\pi$ and $\sigma$ Zeeman components are all displaced from
their zero field positions by different amounts. The quadratic shift is
of similar importance to the linear shift at $B\sim 4$\,MG for the
higher components of the Balmer series (e.g. $H\delta$). The spectra
of NLTT\,12758 indicate that the magnetic component of this system
belongs to the low field regime, as first reported by \citet{kaw2012}.

Before outlining our modelling methods, we need to stress that an
important and as yet unsolved problem regarding the modelling of
magnetic atmospheres, particularly in the high magnetic field regime,
is that concerning line broadening. However in the low field regime,
which is appropriate to the study of the spectrum of NLTT\,12758, it
is possible to assume, as a first approximation, that each Zeeman
component is broadened as in the zero field case. This approach has
been used successfully for the Zeeman modelling of hot white dwarfs
and has allowed the determination of the mass of the hot and
ultra-massive magnetic white dwarfs 1RXS\,J0823.6-2525
\citep[$B\sim 2.8-3.5$\,MG][]{fer1998} and PG\,1658+441
\citep[$B\sim 3.5$\,MG][]{fer1998,sch1992}. In the case of
PG\,1658+441, the spectroscopic mass was found to be in good agreement with 
that determined by the trigonometric parallax method
\citep{dah1999,ven2008}. No trigonometric parallax is as yet
available for 1RXS\,J0823.6-2525. On the other hand, in cool white dwarfs
such as NLTT\,12758, the contribution due to Stark broadening is negligible
and spectral line broadening is dominated by resonance. For H$\alpha$ to
H$\gamma$, we used parameters from the comprehensive self-broadening theory of 
\citet{bar2000}, and for the upper Balmer lines we combined the impact 
parameters from \citet{ali1965,ali1966} with the van der Waals parameters 
as described in \citet{kaw2012b}.

The modelling of the magnetised spectrum of NLTT\,12758 has been
conducted as follows. First, we have computed a zero-field grid of
pure hydrogen white dwarf model atmospheres \citep[see ][]{kaw2012}.
We used the ML2 parameterization of the mixing length theory with
$\alpha=0.6$, where $\alpha$ is the ratio of the mixing length to the
pressure scale height. Convection is predicted to be suppressed in cool
magnetic white dwarfs \citep{tre2015}, however we will investigate the effect 
of suppressed convection on the spectral lines of stars 
such as NLTT~12758 in future work.
This grid of models was then used as input for the
magnetic atmosphere program of \citet{wic1979}, modified to allow for Doppler,
resonance and Stark broadenings and magneto-optical effects which 
take into account the different refractive indices for radiation with different 
polarisation state \citep{mar1981}. The shifts and strengths in hydrogen lines, 
caused by the magnetic field, are included using the results of Zeeman
calculations by \citet{kem1974}.
Atmospheric models were then constructed at selected points on the
visible hemisphere of the white dwarf taking into consideration the
changes in field strength and direction. The resulting Stokes
intensities were then appropriately summed to yield a synthetic
spectrum.

The field geometry is strongly dependent on field strength and
structure and models built on observations obtained at different
phases, if the star rotates around its axis, are better constrained
than those restricted to one single intensity spectrum corresponding
to only one magnetic phase. The best constrained models are those
based on observations at different rotational phases and for which
\emph{both} intensity and polarisation spectra are available as it is
the case for NLTT\,12758.

The modelling of a magnetic white dwarf usually starts with the
assumption that the magnetic field configuration is that of a centred
dipole. Then, if necessary, more complex structures are
investigated. These usually consist of offset dipoles or combinations
of higher order multipoles. For the present set of observations of
NLTT\,12758 we found that a centred dipole model was inadequate to
model the rotationally modulated Zeeman components by simply changing
the viewing angle. This is because a centred dipole allows a field
spread of at most of a factor 2, which is not sufficient to model the
observations of NLTT\,12758.  It is possible to achieve a larger
magnetic field spread by offsetting the dipole from the centre of the
star. If the dipole is shifted by a fraction $a_z$ of the stellar radius
along the dipole axis, then the ratio of the field strengths $B_{p1}$
and $B_{p2}$ at the two opposite poles become
\begin{equation}
\frac{B_{p1}}{B_{p2}}=\left(\frac{1-a_z}{1+a_z}\right)^3
\end{equation}
We describe in detail how we have achieved the best-fit model for
NLTT\,12758 in the sections that follow.

\subsubsection{Spectroscopic analysis}\label{AtmParam}

We fitted the X-shooter spectra with two sets of model spectra. The first
set of model spectra are for non-magnetic hydrogen-rich white dwarfs as 
described in \citet{kaw2012}. The Balmer line profiles used in the
synthetic spectra calculations are described in \citet{kaw2012b}.
The second set of model spectra include a 
magnetic field (as described above).

The procedure fits simultaneously the effective temperature and
surface gravity of both white dwarfs (4 parameters). We used the
mass-radius relations of \citet{ben1999} to scale the flux for both
stars and ensure that the relative flux contribution of each star is
preserved imposing a common distance for both stars. A similar decomposition
method was adopted in the analysis of the hot double degenerate EUVE~J1439+750
\citep{ven1999} and in the analysis of a sample of double degenerates by 
\citet{rol2014} and \citet{rol2015}. The results are model dependent due to
uncertainties in the treatment of line broadening in the presence of a magnetic
field as previously noted by \citet{kul2009}. However, the presence of a 
non-magnetic DA companion with a reliable radius measurement, as in the case 
of NLTT~12758, helps constrain the radius of the magnetic component.
A direct constraint on the stellar radii would be achieved
with a parallax measurement.

The Zeeman
splitting observed in the X-shooter spectra (H$\alpha$ and H$\beta$)
implies an averaged surface field of $B_S = 1.70\pm0.04$\,MG. We used
this value as a starting point to calculate sets of magnetic field
spectra with varying polar field strength and offset. 
We fitted the 
spectra with the following magnetic field
strengths and offsets: offset $= -0.1$ at $B_P = 2.8, 3.1, 3.4, 3.6$
MG; offset $= 0$ at $B_P = 2.6, 3.0, 3.2, 3.4$ MG; offset $= +0.1$ at
$B_P = 2.4, 2.9, 3.1, 3.3$ MG. We also fitted the X-shooter spectra at
viewing angles of $50^\circ$ and $80^\circ$ for each offset and field
strength value. Note that the total exposure time covers nearly two
complete rotation cycles and the viewing angle represents a cycle average.

\begin{figure*}
\includegraphics[width=0.7\textwidth]{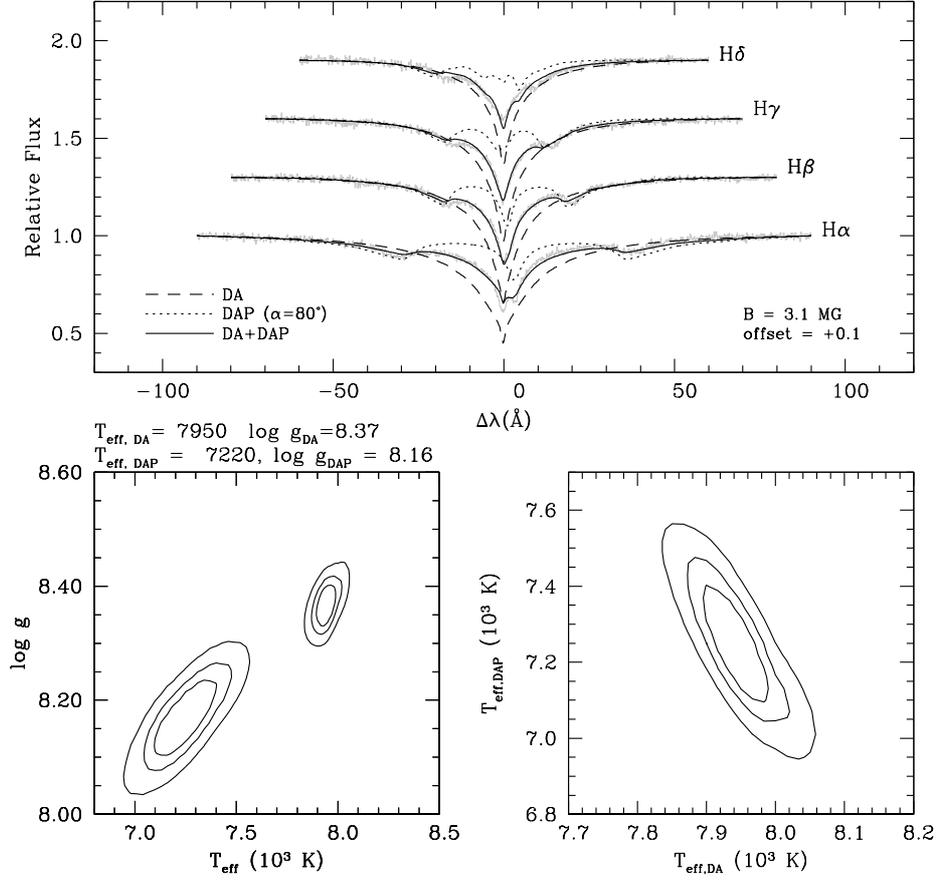}
\caption{(Top panel) observed Balmer line profiles of NLTT\,12758 compared to 
the best-fitting models. The best-fit shows that the components of NLTT\,12758 
are a non-magnetic DA white dwarf (dashed lines) paired with a magnetic DA 
white dwarf (dotted lines). Confidence contours at 66, 90, and 99\% are shown 
in the $T_{\rm eff, DAP}$ vs $T_{\rm eff, DA}$ plane (bottom right) and 
$\log{g}$ vs $T_{\rm eff}$ for both stars (bottom left).}
\label{fig_balmer_fit}
\end{figure*}

Figure~\ref{fig_balmer_fit} compares the X-shooter spectrum and the
best-fitting models for the two stars. The magnetic white dwarf has a
polar magnetic field $B_P =3.1$ MG offset by $a_z = +0.1$ from the
stellar centre. The magnetic white dwarf appears to be slightly
cooler with $T_{\rm eff, DAP} = 7220\pm180$ K and a surface gravity of
$\log{g}=8.16\pm0.08$. The non-magnetic white dwarf is a little hotter and
more massive with $T_{\rm eff, DA} = 7950\pm50$ K and
$\log{g} = 8.37\pm0.04$.  The best-fitting viewing angle to the dipole
axis is on average $\alpha=80^{\circ}$.
Table 4 lists the stellar parameters. We computed the mass and cooling age
of each component using the evolutionary models of \citet{ben1999}. The
spectroscopic mass ratio $M_{\rm DA}/M_{\rm DAP}=1.1-1.3$ is consistent 
with the orbital mass ratio, but also more accurate, and implies that the mass 
of the DA star may be slightly higher than the mass of the DAP star. We 
then estimated the absolute magnitude of each component and calculated the
distance to the system.

\citet{rol2014} and \citet{rol2015} measured the stellar parameters of NLTT\,12758 by
fitting H$\alpha$ together with the spectral energy distribution (SED)
including only $VJHK$. They obtained $T_{\rm eff, DAP} = 6041$ K and
$T_{\rm eff, DA} = 8851$ with a radius ratio of
$R_{\rm DA}/R_{\rm DAP} = 0.908$. Although our radius ratio is in agreement 
with theirs, our effective temperatures differ from their effective 
temperatures.

\begin{figure*}
\includegraphics[viewport=0 365 570 570,clip,width=1.0\textwidth]{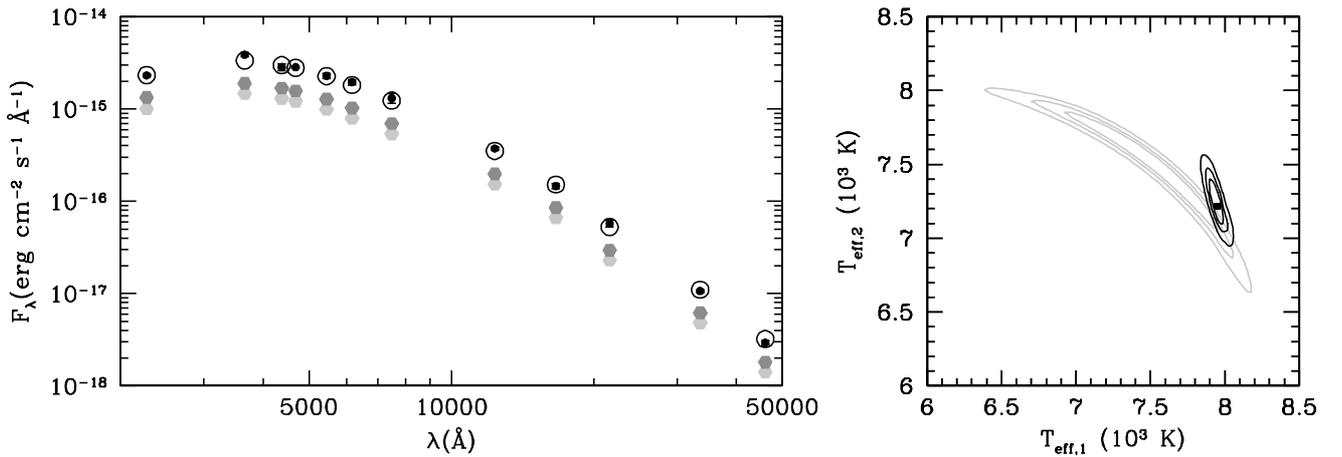}
\caption{The left panel compares the best-fitting photometry (open circle)
to the observed photometry (solid black circles). The contribution of 
individual stars are plotted in different grey shades as hexagonals. The right
panel plots the confidence contours (66, 90, and 99\%) of the spectroscopic 
fit (in black) and the contours of the SED fit (grey full lines). Note that 
$\log{g}=8.4$ for star 1 (DA) and $\log{g}=8.2$ for star 2 (DAP).}
\label{fig_sed_fit}
\end{figure*}

Taking advantage of a broader wavelength coverage, we re-analysed the
SED. First, we fitted the photometric data set
($NUV$,$UBV$,$gri$,$JHK$ and $W1$,$W2$) by fixing the surface gravity
measurements to those obtained in the spectroscopic analysis. We
allowed for both temperatures to vary and assumed null interstellar
extinction. The resulting effective temperatures are nearly in
agreement with the spectroscopic analysis showing that interstellar 
extinction in the line of sight toward NLTT~12758 is negligible when
compared to the total
extinction in the same line of sight, $E(B-V) = 0.06$
\citep{sch1998}. Figure~\ref{fig_sed_fit} shows 
the model photometry fitted to the measured photometry and compares the
confidence contours for the SED fit, as
well as the confidence contours for the Balmer line fit
(Fig.~\ref{fig_balmer_fit}). The overlapping contours show that the two
methods are consistent and imply that the two objects share similar
stellar parameters. In the following discussion we adopt the results
of the spectroscopic analysis.

Our results differ markedly from those of \citet{rol2014} and 
\citet{rol2015} who reported a temperature difference 
$\Delta T=T_{\rm eff, DA}$-$T_{\rm eff, DAP}\approx2800$~K while we estimated 
much closer temperatures for the components ($\Delta T\approx 700$~K). On the 
other hand we estimated a similar mass ratio. Our spectroscopic analysis 
includes the first four members of the Balmer line series (H$\alpha$ to 
H$\delta$), thereby lifting potential degeneracy in the $T_{\rm eff}/\log{g}$ 
solution, while \citet{rol2014} and \citet{rol2015} only include H$\alpha$. 
However, both solutions are model dependent and part of the discrepancy may
also be attributed to different line-broadening prescriptions used in 
calculating magnetic synthetic spectra. The large temperature difference 
reported by \citet{rol2014} and \citet{rol2015} should also be noticeable in 
the SED, particularly in the near ultraviolet. Our own analysis based an 
extensive data set implies a temperature difference no larger than 
$\approx 1100$~K ($1\sigma$) while a larger temperature difference would be 
incompatible with the $GALEX$ $NUV$ measurement.

Using the evolutionary mass-radius relations of \citet{ben1999}, we
find that the cooling ages of the two white dwarfs in NLTT\,12758 are
comparable.  However, \citet{val2014} have proposed that convection in
cool white dwarfs is suppressed by magnetic fields, and therefore
magnetic white dwarfs may appear younger than they are. The 3D
radiation magnetohydrodynamic simulations of \citet{tre2015} have
confirmed that magnetic fields do suppress convection, however they do
not affect the cooling of the white dwarf until temperatures have dropped 
below 6000\,K. Since the magnetic white
dwarf is hotter than this upper limit, it is likely that its age is
not affected and that the two objects formed around the same time.

We derived an orbital inclination of $i = 45^\circ$ for NLTT\,12758 by combining 
the component masses with the orbital parameters
and using:
\begin{equation}
\frac{M_{\rm DA}^3 \sin^3{i}}{(M_{\rm DA}+M_{\rm DAP})^2} = \frac{P K_{\rm DAP}^3}{2\pi G}
\end{equation}
where $M_{\rm DA}$ and $M_{\rm DAP}$ are the masses of the
non-magnetic and magnetic white dwarfs respectively, $P$ is the
orbital period, $K_{\rm DAP}$ is the velocity semi-amplitude of the
magnetic white dwarf and $G$ is the gravitational constant.  Setting
the system inclination at $i=90^\circ$, the maximum orbital period is
$P\approx3.3$~d.

\begin{table}
\centering
\caption{Summary of NLTT\,12758 parameters\label{tbl_param}}
\begin{tabular}{lcc}
\hline
Parameter & DA & DAP \\
\hline
$T_{\rm eff}$ (K)      & $7950\pm50$   & $7220\pm180$ \\
$\log{g}$ (c.g.s)      & $8.37\pm0.04$ & $8.16\pm0.08$ \\
Mass (M$_\odot$)       & $0.83\pm0.03$ & $0.69\pm0.05$ \\
Cooling age (Gyrs)     & $2.2\pm0.2$   & $1.9\pm0.4$ \\
$M_V$ (mag)            & $13.65\pm0.06$& $13.69\pm0.18$ \\
Period (d)             & \multicolumn{2}{c}{$1.15401\pm0.00005$} \\
$K$ (km~s$^{-1}$)      & $81.9\pm17.3$ & $89.7\pm3.8$ \\
$\gamma$ (km~s$^{-1}$) & $94.2\pm17.3$ & $96.4\pm2.6$ \\
$d$ (pc)               & \multicolumn{2}{c}{$32.6\pm3.5$} \\
$v_r$ (km~s$^{-1}$)    & \multicolumn{2}{c}{$58.0\pm3.9$} \\
\hline
\end{tabular}\\
\end{table}

The calculated white dwarf gravitational redshifts ($\gamma_{g,DAP}=38.4\pm2.9$,
$\gamma_{g,DA}=53.6\pm1.7$~km~s$^{-1}$) may be subtracted from their
respective systemic velocities to obtain an estimate of the radial
velocity of the system. Using the more precise velocity of the DAP
star we obtain $v_r=58.0\pm3.9$\,km\,s$^{-1}$.  Combining the proper
motion measurements \citep{kaw2012}, the photometric distance estimate
($d$) and the radial velocity ($v_r$) of the system we determine the
Galactic velocity components
$(U, V, W)=(-40\pm4,-48\pm5,-3\pm6)$~km~s$^{-1}$ which suggest that
the system is relatively young and belongs to the thin disc
\citep{pau2006}.

\subsection{Evolution of NLTT\,12758}\label{evolution}

In order to understand the evolution of NLTT\,12758 we have used the
rapid binary star evolution algorithm, {\sc bse}, of
\citet{Hurley2002}. We have evolved a number of binaries from the Zero
Age Main Sequence (ZAMS) to the age of the Galactic disc \citep[9.5
Gyr, e.g.~][]{Oswalt1996,Liu2000}.  This code is a derivation of the
single star evolution code of \citet{Hurley2000} where the authors use
analytical formulae to approximate the full evolution of stars.  The
{\sc bse} takes into consideration stellar mass-loss, mass transfer,
Roche lobe overflow, CE evolution, tidal interaction,
supernova kicks and angular momentum loss caused by gravitational
radiation and magnetic braking.  In order to model the CE
evolution, the {\sc bse} uses the $\alpha_{\rm CE}$ formalism, where
$\alpha_{\rm CE}$ is a parameter with values in the range $0.1-0.9$.
In our calculations we have adopted $\eta=1.0$ for the Reimers'
mass-loss parameter, as outlined in \citet{bri2015} and a stellar
metallicity, $Z = 0.02$.

We have then generated a synthetic population of binaries with ZAMS
conditions of the mass of the primary star, $M1$, between 3.5 and
4.5\,$M_\odot$, the mass of the secondary star, $M2$, between 2.5 and
3.5\,$M_\odot$ and the initial period $P_0$ in the range
$2000-3500$\,days, as these values were in the region of the expected
initial conditions for the final properties of the components of
NLTT\,12758.  We allowed 200 steps in each parameter in the {\sc bse}
evolution of the population through to the age of the Galactic disk.
In all cases we assumed an initially circular orbit for the progenitor
binary, that is, an eccentricity of zero.  The calculations were
repeated for values of $\alpha_{\rm CE}$ = {\{}0.10, 0.20, 0.25, 0.30,
0.40, 0.50, 0.60, 0.70, 0.80, 0.90{\}}.

A number of stellar types are recognised by {\sc bse} within its logic 
throughout the stages of evolution. These types are set out in Table~1 
of \citet{bri2015}.

The evolved populations were searched for systems that resulted in a
pair of Carbon/Oxygen white dwarfs (CO WDs), that is, type 11s in the
{\sc bse} system.  We have found that as $\alpha_{\rm CE}$ increases
the number of CO WD double degenerate systems increases.  However, not
all of these systems correspond with the type of evolution path that
would lead to the final parameters of NLTT\,12758, i.e. cooling ages,
period, masses.  

A suitable near match to NLTT\,12758 was achieved at
$\alpha_{\rm CE}=0.15$ with initial masses of 3.75\,M$_\odot$ and
2.80\,M$_\odot$ and with an initial period of 2656\,days.  As {\sc bse}
consists of many approximations, the resulting solution is
considered to be satisfactory and within the errors on the
parameters of NLTT\,12758 given in Table\,\ref{tbl_param}.

The evolution shown in Table\,\ref{tab:evol} starts with two stars, S1
and S2, and follows each of them through their normal evolution until
256\,Myr. Up until this time the only interactions between the two
stars are small mass losses due to winds and the consequent small
changes in orbital separation and period. At 256\,Myr the stars start
to interact by common envelope evolution. First, the more massive
star, S1, develops an extended envelope which overflows the Roche
lobe. This draws the stars closer together by friction eroding the
orbit. When Roche lobe overflow ceases and S1 reveals its core as a
CO WD, the two stars are about 588\,R$_\odot$ apart with a period of
around 864\,days. At this point, S2 is still a main sequence star. About
315\,Myr later, S2 initiates its own common envelope evolution
resulting in a second CO WD, an orbital separation of 5.3\,R$_\odot$
and a period of only 1.161\,days. One of the pair, S2, is now a
magnetic WD resulting from the dynamo effect within the common
envelope. S1 loses about 2.8\,M$_\odot$ during the first common
envelope phase while S2 loses about 2\,M$_\odot$. As the second common
envelope evolution brings the two stars very close together by
shrinking the orbit from about 500 to 5\,R$_\odot$, it is S2 that
develops the magnetic field and the rapid rotation.

\begin{table*}
\centering
\caption{Evolution of a binary star system of approximately the size of 
NLTT\,12758 starting from ZAMS through to the end of their interaction 
and the production of a double degenerate WD pair. $M1$ and $M2$ are 
the masses of the primary and secondary stars respectively (in solar 
masses), $S1$ and $S2$ are the stellar types varying throughout their 
evolution as shown in Table 1 in \citet{bri2015}. $Sepn$ is the stellar 
separation in solar radii, $Period$ is the orbital period in days and the 
$Event-Type$ is the event happening to the system at the time given 
in column 2.}
\label{tab:evol}
\begin{tabular}{ r r r r r r r r l}
\hline
\noalign{\smallskip}
 Step & Time  &   M1    &   M2    & S1  & S2 & Period & Sepn & Event-Type\\
        &  (MYr) &(M$_\odot$)&(M$_\odot$)&      &     &   (days)  &(R$_\odot$)\\
\hline
\noalign{\smallskip}
  1 &      0.000 &  3.750 &  2.800 & 1 & 1 &   2656.000  &  1510.578 & ZAMS  \\
  2 &    210.988 &  3.750 &  2.800 & 2 & 1 &   2653.321  &  1509.562 & S1$\Rightarrow$Hertzsprung Gap \\
  3 &    212.057 &  3.750 &  2.800 & 3 & 1 &   2653.713  &  1509.674 & S1$\Rightarrow$RGB  \\
  4 &    212.955 &  3.747 &  2.800 & 4 & 1 &   2655.573  &  1510.204 & S1$\Rightarrow$He core burning \\
  5 &    253.754 &  3.676 &  2.800 & 5 & 1 &   2714.292  &  1526.805 & S1$\Rightarrow$Early AGB \\
  6 &    255.551 &  3.597 &  2.801 & 6 & 1 &   2668.247  &  1503.396 & S1$\Rightarrow$Late AGB \\
  7 &    255.989 &  2.787 &  2.827 & 6 & 1 &   2819.839  &  1493.233 & Begin Roche lobe overflow \\
  8 &    255.989 &  0.827 &  2.827 &11& 1 &    864.356   &   588.342  & CEE, S1$\Rightarrow$CO WD\\
  9 &    255.989 &  0.827 &  2.827 &11& 1 &    864.356   &   588.342  & End Roche lobe overflow \\
 10&    443.089 &  0.827 &  2.827 &11& 1 &    864.356   &   588.342  & S2$\Rightarrow$Blue straggler\\
 11&    449.391 &  0.827 &  2.827 &11& 2 &    864.356   &   588.342  & S2$\Rightarrow$Hertzsprung Gap \\
 12&    452.151 &  0.827 &  2.826 &11& 3 &    864.520   &   588.398  & S2$\Rightarrow$RGB  \\
 13&    455.303 &  0.827 &  2.824 &11& 4 &    865.691   &   588.810  & S2$\Rightarrow$He core burning \\
 14&    567.390 &  0.827 &  2.774 &11& 5 &    889.679   &   596.911  & S2$\Rightarrow$Early AGB \\
 15&    570.808 &  0.828 &  2.725 &11& 6 &    757.173   &   533.592  & S2$\Rightarrow$Late AGB \\
 16&    571.109 &  0.828 &  2.662 &11& 6 &    689.277   &   498.253  & Begin Roche lobe overflow \\
 17&    571.109 &  0.828 &  0.652 &11&11&        1.161   &       5.297  & CEE, S2$\Rightarrow$CO WD \\
 18&    571.109 &  0.828 &  0.652 &11&11&        1.161   &       5.297  & End Roche lobe overflow \\
 19&   2791.209&  0.828 &  0.652 &11&11&        1.154   &       5.278  & Present Day \\
\noalign{\smallskip}
\hline
\end{tabular}
\end{table*}

From this time the pair interact by gravitational radiation and
magnetic braking with consequent orbital shrinkage until at 2791\,Myr
they reach the present day with a separation of 5.278\,R$_\odot$ and
an orbital period of 1.154\,days.  The cooling ages are 2535\,Myr and
2220\,Myr for the non-magnetic and magnetic white dwarfs respectively
\citep[for details of the method see ][]{bri2015}. Further evolution will see the orbit shrinking further
until at some stage (over a time much longer than a Hubble time) Roche
lobe overflow restarts and the two stars merge. The possible final
fate of double degenerate white dwarf systems, such as NLTT\,12758, is
discussed in the section below.

\section{Discussion}

NLTT\,12758 is a member of a growing class of double degenerate
systems consisting of two white dwarfs, one magnetic and one
not. Table~\ref{tbl_dd} lists the currently known double degenerate
systems containing at least one magnetic white dwarf. The table lists
the names, orbital and rotational periods, the magnetic field
strength, effective temperatures and masses of the components. It
includes both close binaries and CPM systems. Most of the systems for
which effective temperatures and masses are determined appear to have
formed, within uncertainties, at the same time. In the case of CPM
systems, where it is assumed that the stars did not interact during
their evolution, there are systems with inconsistencies in their ages
if one assumes single star evolution for each star. Apart from the
well documented case of EUVE\,J0317-855 \citep{fer1997}, another more
recent example is given by PG\,1258+593 and its common proper motion
magnetic white dwarf companion SDSS\,J1300+5904.  \citet{gir2010}
found that the masses of these white dwarfs are
0.54$\pm$0.06\,M$_\odot$ for the non-magnetic and
0.54$\pm$0.01\,M$_\odot$ for the magnetic component.  Despite their
very similar masses, SDSS\,J1300+5904 is a cool white dwarf
($T_{\rm eff}=6300\pm300$\,K) while PG\,1258+593 is substantially
hotter ($T_{\rm eff}=14790\pm77$\,K). \citet{gir2010} find that the
temperature discrepancy gives a difference in cooling age (and thus in
formation age of the white dwarfs) of $1.67\pm 0.05$\,Gyr. If one
makes the plausible assumption that the progenitors of these CPM white
dwarfs formed in the same protostellar cloud at roughly the same time,
then the similar white dwarf masses and their large age discrepancy
give rise to a paradox. A possible solution is that this system was
initially a triple system where two stars interacted and merged to
form the magnetic white dwarf SDSS\,J1300+5904 about 1.67\,Gyr before
the third non-interacting object evolved into the non-magnetic white
dwarf PG\,1258+593.

However, the situation appears to be rather different for the double
degenerate system NLTT\,12758, as reported in section
\ref{evolution}. Since NLTT\,12758 is a close binary system, it is
highly unlikely that the field of the magnetic component was caused by
the merging of two stars in an initially triple system. Instead, the
magnetic field must have originated during CE evolution
in a manner very similar to that occurring during the formation of a
magnetic cataclysmic variable, as proposed by \citet{bri2016} (in
preparation).  In this scenario, the closer the cores of the two stars
are drawn during CE evolution, the greater the
differential rotation and thus the larger the dynamo generated field
will be. If CE evolution leads to the merging of the two
stellar cores the resulting object would be an isolated highly
magnetic white dwarf \cite[see][]{wic2014}. If the two stars do not
coalesce they are expected to emerge from the CE as close
binaries that are already interacting, and thus appear as magnetic
cataclysmic variables, or are close to interaction. The low-accretion
rate polars, where a magnetic white dwarf accretes matter from its
companion through a stellar wind, have been suggested by
\citet{sch2009} to be the progenitors of the polars, which are the
highest field magnetic cataclysmic variables. In the polars a magnetic
white dwarf accretes matter from an un-evolved low-mass (M-dwarf)
companion via magnetically confined accretion flows. The orbital
periods are typically between 70 minutes to a few hours and Zeeman and
cyclotron spectroscopy from the UV to the IR bands have revealed the
presence of fields between 7 to 230\,MG
\citep[e.g. see][]{fer1992,fer1993,fer1996,sch2001b} in the case of the 
polars, and 1 to 20\,MG in the case of intermediate polars \citep{fwk1993}. 
The difference between these systems and NLTT\,12758 is that both progenitor 
stars of NLTT\,12758 were too massive to evolve into a magnetic cataclysmic
variable. However, the indications
seem to be that the magnetic white dwarf component of NLTT\,12758
acquired its field via a mechanism similar to that propounded to
explain the origin of magnetic cataclysmic variables.

The properties of NLTT~12758 mean that the two white dwarfs 
will coalesce in a time much longer than a Hubble time
\citep[$\sim 140$\,Gyr;][]{rit1986}, however it is still interesting to
speculate what the final fate of a system like this might be.

The first simulations of two merging CO WDs were conducted by
\citet{sai1985} and showed that the fast mass accretion rate
($\gtrsim 10^{-5}$\,M$_\odot$yr$^{-1}$) from the less massive to the
more massive white dwarf ignites an off-centre carbon flash. The
carbon nuclear burning then propagates toward the stellar centre
turning the CO WD into an ONe WD quiescently. The outcome of such an
event would not be a carbon deflagration but an accretion induced
collapse (AIC) triggered by electron captures on $^{24}$Mg and
$^{20}$Ne. The result would be a rapidly spinning neutron star that
would appear as an isolated millisecond pulsar \citep[MSP,
e.g.][]{lor2008}. The low space velocities of isolated MSPs suggest
that there could not have been a substantial SNII kick imparted to the
emerging neutron star, thus supporting the AIC hypothesis
\citep{fer2007,hur2010}. The calculations of \citet{chen2013} lend
further support to this idea since they show that it is unlikely that
the isolated MSPs may be generated via the LMXB recycling scenario because
this would require the total ablation of their donor star. Thus, merging
events of systems similar to NLTT\,12758, but with initial
parameters that would allow faster evolutionary timescales, could
provide a simple explanation for the existence of isolated MSPs. 

On the other hand, the merging of the two stars in NLTT\,12758 may
give rise to a supernova event. Recent simulations conducted by
\citet{dan2014} and \citet{dan2015} showed that a merging system with
a total mass M$_{\rm tot} \ge 2.1$\,M$_\odot$ and comprised of two
white dwarfs of similar mass may result in a Type Ia supernova; The
total mass of NLTT\,12758, $M_{\rm tot} = 1.52$\,M$_\odot$, would be
below the predicted cutoff for this event to occur. However,
other studies conducted by \citet{pak2011} and \citet{sat2016} found
that systems with a mass ratio greater than $\sim 0.8$ could indeed
result into a SNIa explosion.  Clearly, a consensus in this area of
research still needs to be reached \citep[e.g.][]{fer2013}.

\begin{table*}
\centering
\caption{Known double degenerates containing a magnetic white dwarf\label{tbl_dd}}
\begin{tabular}{llcccccccc}
\hline
Name & Alternate name & $P_{\rm orb}$  & $P_{\rm rot}$  & $B$  & \multicolumn{2}{c}{$T_{\rm eff}$ (K)} & \multicolumn{2}{c}{Mass ($M_\odot$)} & Reference \\
     &                &                &                & (MG) &  Magnetic        &   Companion        &   Magnetic         & Companion       &           \\
\hline
0040+000 & SDSS~J004248.19+001955.3 & ...    & ...    & 14         & \multicolumn{2}{c}{11000}   & ...           & ...           & 1 \\
0121-429$^a$ & LHS 1243             & ...    & ...    & 10.3       & 6105         & 5833         & 0.7$^d$       & 0.54$^d$      & 2,3,4 \\
0239+109$^a$ & G 4-34, LTT 10886    & ...    & ...    & 0.7        & 10060        & 7620         & ...           & ...           & 5,6 \\
0325-857 & EUVE~J0317-855           & $\sim2095$ yr & 725 s & 185-425 & 33000     & 16360        & 1.3           & 0.85          & 7,8,9,10 \\
0410-114 & NLTT\,12758, G160-51      & 1.15 d & 23 min & 3.1       & 7220         & 7950         & 0.69          & 0.83          & This work \\
0512+284$^a$ & LSPM~J0515+2839      & ...    & ...    & 2.15       & 5940         & 6167         & 0.81$^d$      & 0.61$^d$      & 3,4 \\
0745+303 & SDSS~J074853.07+302543.5 & CPM    & ...    & 11.4       & 21000        & 22702        & 0.81          & 0.88          & 11 \\
0843+488$^b$ & SDSS~J084716.21+484220.4 & ...& ...    &            & \multicolumn{2}{c}{19000}   & ...           & ...           & 1 \\
0924+135 & SDSS~J092646.88+132134.5 & CPM    & ...    & 210        & 9500         & 10482        & 0.62          & 0.79          & 12 \\
0945+246 & LB~11146                 & $\sim130$ d &...& $\sim 670$ & 16000        & 14500        & 0.90          & 0.91          & 13,14,15 \\
1026+117$^a$ & LHS~2273             & ...    & ...    & 17.8       & 5691         & 7350         & 0.75$^d$      & 0.64$^d$      & 3,4 \\
1258+593 & SDSS~J130033.48+590407.0 & CPM    & ...    & 6          & 6300         & 14790        & 0.54          & 0.54          & 16 \\
1330+015$^a$ & G~62-46              & ...    & ...    & 7.4        & 5712         & 7618         & 0.82$^d$      & 0.64$^d$      & 3,4 \\
1440+753 & EUVE~J1439+750           & ...    & ...    & 10         & 42000        & 30000        & 0.9           & 1.1           & 17 \\
1503-070$^a$ & GD~175               & ...    & ...    & 2.9        & 6062         & 7051         & 0.95$^d$      & 0.73$^d$      & 3,4 \\
1506+399 & CBS~229                  & CPM    & ...    & 18.9       & 18000        & 16761        & 0.81          & 0.82          & 11 \\
1506+523 & SDSS~J150746.80+520958.0 & CPM    & ...    & 65.2       & 18000        & 17622        & 0.99          & 0.70          & 12 \\
1514+282$^a$ & SDSS~J151625.07+280320.9 &... & ...    & 2.05       & 7168         & 7662         & 0.77$^d$      & 0.54$^d$      & 3,4 \\
1713+393$^a$ & NLTT~44447           & ...    & ...    & 2.1        & 6204         & 6556         & 0.94$^d$      & 0.54$^d$      & 3,4,18 \\
1814+248$^c$ & G~183-35             & ...    & ...    & 12.05/7.8  & 5998         & 5849         & 0.85$^d$      & 0.74$^d$      & 3,4,19 \\
1818+126$^a$ & G~141-2              & ...    & ...    & 3.75       & 5215         & 6451         & 0.64$^d$      & 0.54$^d$      & 3,4 \\
\hline
\end{tabular}\\
\flushleft
$^a$ DAH+DC\\
$^b$ DAH+DB\\
$^c$ DAH+DAH\\
$^d$ Masses are calculated using the mass-radius relations of \citet{ben1999}, 
the published parameters of the magnetic star and ratio of the stellar radii. \\
References: (1) \citet{sch2003}; (2) \citet{sub2007}; (3) \citet{rol2014}; (4) \citet{rol2015}; (5) \citet{koe2009}; (6) \citet{gia2011};
(7) \citet{fer1997}; (8) \citet{ven2003}; (9) \citet{kul2010}; (10) \citet{law2013}; (11) \citet{dob2013}; (12) \citet{dob2012};
(13) \citet{nel2007}; (14) \citet{gle1994}; (15) \citet{lie1993}; (16) \citet{gir2010}; (17) \citet{ven1999}; (18) \citet{kaw2006};
(19) \citet{put1995}
\end{table*}

\section{Conclusions}

In this paper we have reported our studies on the close,
super-Chandrasekhar double degenerate system NLTT\,12758 consisting of
two CO WDs of similar masses and ages and with one of the two
components highly magnetic. The magnetic white dwarf spins around its
axis with a period of 23 minutes and they orbit around each other with
a period of 1.15 days. Although the components of NLTT\,12758 will not merge
over a Hubble time, systems with very similar initial parameters will
come into contact and merge thus undergoing either an accretion
induced collapse to become a rapidly spinning neutron star (an
isolated MSP) or a Type Ia supernova explosion. Given the
theoretical uncertainties, the jury is still out on the fate of such
systems.

\section*{Acknowledgements}

AK and SV acknowledge support from the Grant Agency of the Czech
Republic (P209/12/0217 and 15-15943S) and the Ministry of Education of 
the Czech Republic (LG14013). This work was also supported by
the project RVO:67985815 in the Czech Republic.  SV acknowledges
support from the Mathematical Sciences Institute of the Australian
National University. EP acknowledges support by the Ministry of Education 
of the Czech Republic (grant LG15010).
GPB gratefully acknowledges receipt of an Australian Postgraduate
Award. We thank the referee, Pier-Emmanuel Tremblay, for a thorough report 
and helpful comments on line-broadening theory.

\label{lastpage}

\end{document}